\documentclass[letterpaper,11pt]{article}
\usepackage{chicago}
\usepackage{verbatim}
\usepackage{xspace}
\usepackage{amsmath}
\usepackage{amsthm}
\usepackage{amssymb}
\usepackage{graphicx}
\usepackage{epstopdf}
\usepackage{url}
\newcommand{\commentout}[1]{}

\title{Is state-dependent valuation more adaptive than simpler rules?}
\author{Joseph Y. Halpern\\
Cornell University\\
halpern@cs.cornell.edu 
\and Lior Seeman\\
Uber\\
lior.seeman@gmail.com}
\date{}

\newtheorem*{theorem*}{Theorem}
\newtheorem*{corollary*}{Corollary}
\newtheorem*{conjecture*}{Conjecture}
\newtheorem*{lemma*}{Lemma}
\newtheorem*{thm*}{Theorem}
\newtheorem*{prop*}{Proposition}
\newtheorem*{obs*}{Observation}
\newtheorem*{rem*}{Remark}
\newtheorem*{definition*}{Definition}
\newtheorem*{exm*}{Example}

\begin{document}

\maketitle

\bigskip

\begin{abstract}
McNamara, Trimmer, and Houston \citeyear{MTH12} claim to provide an
explanation of certain systematic deviations from rational behavior
using a mechanism that could arise through natural selection.   
We provide an arguably much simpler mechanism
in terms of computational limitations,
that performs better in the environment described by McNamara, Trimmer, and Houston \citeyear{MTH12}.
To argue convincingly
that animals' use of state-dependent valuation is adaptive
and is likely to be selected for by natural selection,
one must argue that, in some
sense, it is a better approach than the simple strategies that we
propose.
\end{abstract}

\vfill
\paragraph{Acknowledgments:}
This
work was supported in part by NSF grants
IIS-0911036 and CCF-1214844, by	ARO grant W911NF-14-1-0017, 
by Simons Foundation grant \#315783,
and by the
Multidisciplinary University Research Initiative (MURI) program
administered by the AFOSR under grant FA9550-12-1-0040.
%
Thanks to Arnon Lotem,
Alex Kacelnik, 
and Pete Trimmer
 for 
their 
 useful comments.
 Most of this work was carried out while the second author was at
 Harvard's Center for Research on Computation and Society; Harvard's
 support is gratefully acknowledged. 

\eject

\commentout{
%


\noindent {\bf Lay Summary:} McNamara, Trimmer, and Houston claim to
provide an explanation of certain systematic deviations from rational
behavior. We provide an arguably much simpler explanation in terms of
computational limitations.  To argue
convincingly that McNamara et al.'s mechanism is adaptive, and is
likely to be selected for by natural selection, one must argue that,
in some sense, it is a better approach than the simple strategies that we
propose. 

\medskip
\noindent {\bf Keywords:} natural selection; learning; decision
making; bounded rationality 
}

%


\pagenumbering{arabic}
\section{Introduction}\label{sec:intro}

Although much animal behavior can be understood as rational, in the
sense of making a best response in all situations, some systematic
deviations from rationality have been observed.  For example, Marsh,
Schuck-Paim, and Kacelnik \citeyear{MSK04} 
presented starlings with two potential food sources, one which had provided food
during ``tough times'', when the birds had been kept at
low weight, while other had provided food during ``good times'', when
the birds were well fed.  They showed that the starlings preferred the
food source that had fed them during the tough times, even when that
source had a longer delay to food than the other source.  Similar
behavior was also observed in fish and desert locusts
\cite{AHPK09,PKB06}.  

McNamara, Trimmer, and Houston \citeyear{MTH12} claim to provide an
explanation of this behavior using a mechanism that could arise
through natural selection.  They provide an abstract model of the
bird-feeding setting where a decision maker can choose either a
``risky'' action or a ``safe'' action.  They also provide a mechanism
that takes  
internal state into account 
and can lead to good results (where, in
the example above, the internal state could include the
fitness of each source).  However, as we observe, for the particular
parameters used in their model, there is a \emph{much} 
better (in the sense of getting a higher survival probability) and much simpler
approach than their mechanism that does not take the internal
state into account: simply playing safe all the time. It is hard to
see how the mechanism proposed by McNamara et al.~could arise in the
model that they use by natural selection; the simpler mechanism would
almost surely arise instead.  

The fact that always playing safe does well depends on the particular
parameter settings used  by McNamara et al.  Playing safe would not be
a good idea for other parameter settings.  However, we show that a 
simple $2$-state automaton that more or less plays according to what
it last got also does quite well.  It does significantly better
than the McNamara et al.~mechanism, and does well in a wide variety of
settings.   Although our automaton also takes internal state into
account (the internal state keeps track of the payoff at the last
step), it does so in a minimal way, which does not suffice to explain
the irrational behavior observed.


It seems to us that to argue convincingly that the type of mechanism
proposed by McNamara et al.~is adaptive, 
and is likely to be selected
for by natural selection, 
and thus explains animals' use of state-dependent valuation,
then one must argue that, in some
sense, it is a better approach than the simple strategies that we propose.
Now it could be that the simple strategies we consider
do not work so well in a
somewhat more complicated setting, and in that setting, taking the
McNamara et al.'s approach does indeed do better.  
However, such a setting should be demonstrated; it does not seem
easy to do so.   In any case, at a minimum, these observations suggest
that McNamara et al.'s explanation for the use of state-dependent
strategies is incomplete.

We should add that we are very sympathetic to the general approach
taken by McNamara et al., although our motivation has come more from
the work of Wilson \citeyear{W02} and Halpern, Pass, and Seeman
\citeyear{HPS12,HPS13}, which tries to 
explain seemingly irrational
behavior, this time on the part of humans, in an appropriate model.
That work assumes that people are resource-bounded, which is
captured by modeling people as finite-state automata,
and argues that an optimal (or
close to optimal) finite-state automaton will exhibit some of the
``irrational'' behavior that we observe in people.
(The $2$-state automaton that we mentioned above is in fact a special
case of a more general family of automata 
considered in \cite{HPS12}; see
Section~\ref{sec:automaton}.)
We believe that taking computational limitations seriously might be a
useful approach in understanding animal behavior, and may explain at
least some apparently irrational behavior.


The rest of this paper is organized as follows.  
In Section~\ref{sec:model}, we review the model used by McNamara et
al.~\citeyear{MTH12} and compare it to that of \cite{HPS12}.   
In Section~\ref{sec:strategies}, we describe four strategies that an
agent can use in the McNamara et al.~model,
under the assumption that the agent knows which action is the risky
action and which is the safe action.
One is the strategy used
by McNamara et al.; another is a simplification of the strategy that
we considered in our work; the remaining two are baseline strategies.
In Section~\ref{sec:evaluation}, we evaluate the strategies under
various settings of the model parameters.
In Section~\ref{sec:learning}, we consider what happens if the agent
does not know which action is risky and which is safe and, more
generally, the issue of learning.
We conclude in
Section~\ref{sec:discussion}.

\section{The model}\label{sec:model}
McNamara et al. \citeyear{MTH12} assume that agents live at most
one year, and that each year is divided into two periods, winter and
summer. Animals can starve to death during a winter if they do not
find enough food.  If an agent survives the winter, then it
reproduces over the summer, and reproductive success is independent of
the winter behavior.  


A ``winter" is a series of $T$ discrete time steps. At any given time,
the environment is in one of two states: $G$ (good)  or $S$ (sparse);
the state of 
the environment is hidden from the agent.
At every time step there is a small probability $z$ of the environment
switching states.
At each time step, there are two actions potentially available to the
agent, $A$ (which we
think of as the ``risky'' action)
or $B$ (the ``safe'' action).
(The names ``safe'' and ``risky'' are due to the fact 
that the reward swings, depending on whether the environment is good
or scarce, are greater for $A$ than for $B$.)
With probability $\gamma$, both options are available to the
agent; with probability $(1-\gamma)/
2$, the agent must play $A$; and
with probability $(1-\gamma)/2$, the agent must play $B$.
The payoff of actions $A$ and $B$ depends on whether the state of the
environment is $G$ or $S$.

An agent has a certain level of ``energy reserves", denoted by an
integer between 0 and 10.  The maximum level of energy reserves is
thus 10; an agent dies if his energy-reserve level is $0$. At each time
step, one unit of energy reserves is consumed. At each time step, an
agent receives $0,1$ 
or $2$ units of energy. The probability of each of these amounts is
drawn from a binomial distribution $bin(2,p)$
(so that the probability of receiving 0 units is $(1-p)^2$, the probability
of receiving 1 unit is $2p(1-p)$, and the probability of receiving 2
units is $p^2$), 
where $p$ depends
on the current environment state and the choice of action. $P_{GA}$ is used
to denote the probability $p$ when the environment is in state $G$ and the
agent plays action $A$; we can similarly define $P_{GB}$, $P_{SA}$,
and $P_{SB}$.  McNamara et al. assume that rewards are higher in
expectation in the good environment for both actions, that is,  $P_{GA}\geq
P_{SA}$ and $P_{GB}\geq P_{SB}$); moreover, $A$ is the better action in
the good environment,  while $B$ is better in
the sparse environment, so $P_{GA}\geq P_{GB}$ and $P_{SB}\geq P_{SA}$.

It is interesting to compare this model to that used by Halpern, Pass,
and Seeman \citeyear{HPS12}.
Although,  we mentioned in the introduction, their goal was to study
irrational behavior in humans, and the kinds of behaviors
considered were quite different from those considered by McNamara
et al. (the focus was on modeling the behavior in game playing
reported by Erev, Ert, and Roth \citeyear{erev2010entry}), the models
are surprisingly similar. 
The main differences between the two models is 
that, in the model of \cite{HPS12}, an agent's objective is to
maximize his expected 
average payoff over rounds (rather than just to maximize the
probability of surviving for a year).  Agents never
die; 
and an agent's utility is taken to be the limit of his average reward per
round over an infinite time horizon.   
In the language of McNamara, Trimmer, and Houston,  Halpern,
Pass, and Seeman take $\gamma=1$, 
so that both actions are always available.
Moreover, instead of observing the payoff (which is deterministically
dependent on the state of the world), an agent gets a signal
correlated with the real state of the environment when he plays $A$,
and no signal when he plays $B$. As discussed by Halpern et
al. \citeyear{HPS12}, getting a noisy payoff as in the McNamara et
al.~\citeyear{MTH12} model has essentially the same effect as getting a signal
correlated with the environment's state.
As we discuss later, in the scenarios we study here, $P_{GB}$ and
$P_{SB}$ are very close and thus the signal we get about the
environment by playing $B$ is very weak in this model as well.  

\section{Four Strategies}\label{sec:strategies}
In this section, 
we describe four strategies that an agent can use in the McNamara et
al.~model. 
We will be
interested in the probability that an agent survives 
a ``winter" period using each of these strategies.
 Note that the higher this probability is, the
greater  the probability that this strategy will emerge as the dominant
strategy in an evolutionary process. 

\subsection{Baseline strategies}
We consider two baseline strategies. The first is called the \emph{oracle
strategy}.  With this strategy, we assume that an agent knows the true
state of the environment 
before choosing his action, and thus can make the optimal choice in
each round. While this strategy cannot be implemented by an agent  in
this model, we use it to provide an upper bound on the 
the survival probability of the agent.  Clearly, no strategy can do
better than the oracle strategy.

The second baseline strategy we consider is the \emph{safe
  strategy}. With the safe
strategy, the agent always  plays the safe action ($B$) when that choice is
available (recall that in some rounds the agent is forced to
play $A$). 

\subsection{The value strategy}
The strategy studied by McNamara et al.~\citeyear{MTH12}, which we call
the \emph{value strategy}, is based on keeping a value $V(\cdot)$ for
each of the actions and choosing the action with the highest value in
each round. This value is updated in every round using the formula
$V_{new}=(1-\beta)V_{old}+\beta w$, where $\beta$ is a fixed parameter
controlling the learning rate and $w$ is the \emph{perceived reward},
which is defined in more detail below.

In a little more detail, $V(i)$ is initialized to the expected
energy reward of action $i$.  In a round where action $i$ is performed,
$V(i)$ is updated using the formula above (taking $V_{old}(i)$ to
be the currently stored value of $V(i)$ and $V_{new}(i)$ to be the
updated value), where
$w=ue^{k(r-5)}$, $u$ is the number of energy units received 
as a result of performing action $i$, 
$r$ is the current reserve level, and $k$ is a fixed constant
that might evolve to match the scenario parameters. 

In a round where the agent can play
both $A$ and $B$ (which will be the case with probability $1-\gamma$),
it plays whichever one has higher value.  That is, it plays $A$ if 
$V(A) \ge V(B)$, and otherwise plays $B$.

\subsection{The automaton strategy}\label{sec:automaton}

The last strategy we consider is inspired by the strategy used by
Halpern, Pass, and Seeman~\citeyear{HPS12}; we call it the \emph{automaton
strategy}.  
With the automaton strategy, an agent keeps an internal state
that is correlated with the number of good and bad signals it has seen
recently.  Thus, the internal state is not determined by 
the agent's internal reserves, but rather by 
recent observations. 
This strategy is described by a finite automaton with $n$ states
denoted  $[0,\ldots,n-1]$.  If the automaton has a choice, then it
plays action $B$ (the
``safe'' action) in state 0 and plays action $A$ in all the remaining
states.  (Thus, if the automaton has only one state, it plays the safe
strategy.)  The automaton changes state depending on the signal it observes.
Halpern, Pass, and Seeman 
assumed that an automaton in state $0 \le i < n$ 
moved to $i+1$ with probability $p_{up}$
whenever it received a signal that was highly correlated with the
environment being in state $G$; an automaton in state $i > 0$ moved to  
state to $i-1$ with probability $p_{down}$ whenever it 
receives a signal that was highly correlated with the environment
being in state $S$; and an automaton in state 0 (which never received
a signal that would cause it to move in our earlier work, since it
always played $B$) moved to state 1 with a small ``exploration''
probability $p_{exp}$.  If an automaton did not change state according
to the rules above, it just remains  in the same state.

For the purposes of this paper, we take ``received a signal highly correlated
with environment being in state $G$'' to mean ``played $A$ and got a
reward of 2''; while ``received a signal highly correlated with the
environment being in state $S$'' means ``played $A$ and got a
reward of 0''.  (We justify these choices in the next section.)
For simplicity, we take $p_{up} = p_{down} = 1$ and
$p_{exp} = 0$, so that the automaton is deterministic.  Note that we
are able to take $p_{exp} = 0$ because 
$\gamma < 1$, 
so that even in
state 0, the automaton will play $A$ with probability $(1-\gamma)/2$.
(McNamara et al.~\citeyear{MTH12} also point out that randomization is
useful when $\gamma=1$.)


\section{Evaluating the Strategies}\label{sec:evaluation}

In this section we evaluate the four strategies discussed in the previous
section under various settings of the
model parameters. We calculate the survival probability of an agent
using the strategy over a winter of length $T=500$ steps by simulating
$100000$ winters and looking at the ratio of runs in which the agent
survived. We initialize the environment to the sparse state and the
resource level to the maximum of $10$. 

\subsection{Using the McNamara et al.~parameter settings}

We first study the performance of these strategies using 
the baseline parameter
setting considered by McNamara et al.~\citeyear{MTH12}: 
the environment changes with probability $z=0.02$ at each round; the
payoffs are
$P_{GA}=0.9,P_{GB}=0.7,P_{SA}=0.4,P_{SB}=0.6$; and $\gamma=0.5$, so half
the time, an agent has both options available, one quarter of the
time, $A$ must be played, and one quarter of the time, $B$ must be played.
Note that with these parameter settings, when playing $B$, the
expected gain in energy reserves is very close if the environment is
in state $S$ and in state $G$.  Thus, the agent does not learn much
about the state of the environment when playing $B$.  Thus, these
parameter settings lead to a situation similar to that in
\cite{HPS12}, where it is assumed that the agent gets no information when it plays $B$. 
On the other hand, there is a significant gap in the
expected gain in energy reserves between $P_{SA}$ and $P_{GA}$,
so the agent can make useful inferences about the state of the
environment based on its reward when it plays $A$.

According to McNamara et
al, these parameter settings were chosen to ensure a high survival
probability. Indeed, our baseline oracle strategy has a 
survival probability of roughly $91\%$.  
With McNamara et al.'s parameter settings, an animal has an
expected positive gain of energy resources 
in each round, even though it is forced to play $B$ one quarter of the
time. In the good environment it gets an even larger expected gain in
energy resources and thus gets to its maximum level with high
probability. 
Thus, its survival probability is high, although it might still die even with this strategy.

With these parameter settings, McNamara et al.~found that the
value strategy performed best with the constant $k$ used in evaluating $w$ set
to $-.2$, so we used this value in our
experiments; we also follow McNamara et al.~in taking the parameter
$\beta$ to be  $.02$.   
With these settings, the value strategy has a survival probability
of just below $80\%$
(which matches the findings in McNamara et al.~\citeyear{MTH12}). 
Note that the negative value of $k$ says that
the strategy gives greater perceived value for a reward received when its
resource are low.   
McNamara et al.~view this as a justification for animals preferring
option $B$.  

Although the value strategy does well, the safe strategy does even
better.  Indeed, its survival rate is almost 91\%, that is, just about
as good as that of the oracle strategy.  
This is actually straightforward to check analytically as well.%
\footnote{We can view the situation as a Markov chain
with 21 states: $0,
(1,G), (1,S), \ldots, (10,G), (10,S)$, where a state 
$(i,x)$ says that the agent has $i$ units of energy reserves and the
environment is in state $x$; we identify $(0,G)$ and $(0,S)$ since the
agent is dead in either case.  Given a strategy, 
we can easily calculate the transition probabilities between states.
We get a $21 \times 21$ matrix $M$ that describes the one-step
transition probabilities.  Thus, $M^T$ (the result of multiplying $M$
by itself $T$ times) describes the probabilities associated with a
$T$-step transition.   For $T=500$, the choice made by McNamara et
al., and the transition probabilities determined by the safe strategy,
we can compute $M^{500}$ quickly (using standard computer science
algorithms involving repeated squaring of the matrix).  We can read
off the probability $p$ of 
starting in state $(10,S)$ and getting to state $0$ from the matrix;
this is the probability of dying.  Then  $1-p$ is the survival probability.}
It is not
hard to explain the good performance of the safe strategy
compared to the oracle strategy.
In the sparse environment it plays exactly as the oracle strategy, 
while in the good environment, since the expected return at each step
even when playing $B$ is high, it reaches the 
limit of $10$ units with high 
probability by playing $B$; playing $A$ has no real advantage. 

For the same reasons, 
the automaton strategy also performs significantly better than the
value strategy. An
automaton with only 2 states and with deterministic transitions (so
$p_{up}=p_{down}=1$) has a survival probability of over 85\%.
Adding more states to the automaton did not help with this setting of
the model parameters.

\begin{figure}
                \centering
                \includegraphics[width=1\textwidth]{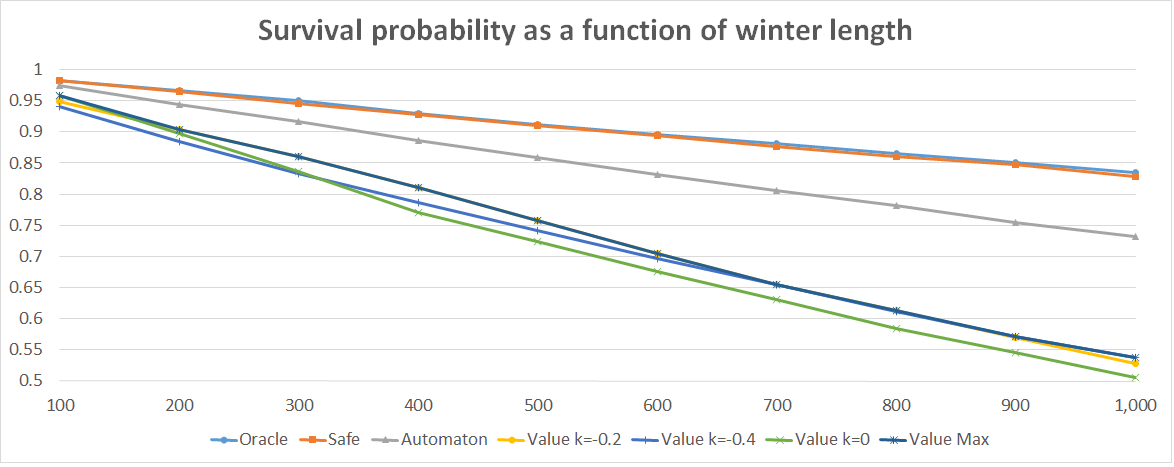} 
 	\caption{Survival probability with different winter lengths}
	\label{fig:lengths}
\end{figure}

We tested the strategies with different lengths of
winter, ranging from $100$ to $1000$ steps, to see how robust these
outcomes are.  For
the value strategy we tested $3$ different values of $k$, since
McNamara et al.~\citeyear{MTH12} found that different optimal values
evolve for different winter lengths. We compare the performance of all
three, as well as just the maximum of them with all other
strategies. As can be seen in Figure~\ref{fig:lengths}, the
safe strategy consistently perform as well as the oracle
strategy. Also, the automaton strategy performs consistently
better than the value strategy, even with only 2 states and compared
to the best choice of $k$ for each choice of $T$, 
and the gap increases as $T$ grows. 

The reason that the value strategy underperforms in these settings is that
although there is no real advantage to playing $A$, after a long period of
the environment being in state $G$, the value strategy sticks with 
$A$ for a while, which puts it at risk when the environment
switches to $B$. (The same is true for automata with extra states.) 

\subsection{Being safe is not always better}
While the safe strategy
performed just about as well as the optimal oracle strategy for the
parameter settings used by McNamara et al., it did not do as well for
other parameter settings.  Recall that the safe strategy is just
the automaton strategy with one state, 
which can be viewed as saying
that while an automaton can get away with being ``dumb'' for the
parameter settings considered by McNamara et al., there exist
environment conditions where more states are desirable. 
In particular, this happens when the expected payoff of playing $B$
drops and $\gamma$ gets closer to $1$, so that the agent is not forced
to play $A$ in many rounds.

For example, when $P_{GA}=0.8, P_{GB}=0.55, P_{SA}=0.3, P_{SB}=0.55$
(note that the expected value of both actions is the same) and
$\gamma=0.9$ we get that the survival probability of the automaton
strategy with $2$ states is $\sim\, 74\%$ while the safe strategy only
gives $\sim\, 70\%$
(the oracle strategy has $\sim\, 86\%$ survival probability). 
This gap grows as $T$ grows. The value
strategy performs quite badly with these parameter settings; it leads
to a survival probability of only $\sim\,  5\%$ (for the 
best choice of $k$).

It is not hard to explain these results. In these settings, the reward
from playing $B$ is not enough to guarantee survival with high
probability, so the automaton strategy gains from being able to play
$A$. In addition, the reward from $A$ is a very good signal of the
state of the environment, so the automaton strategy does not keep
playing $A$ for too long when it should not. The value
strategy does not do so well in this case, since playing $A$ in the
sparse environment is quite bad, while when the environment is in a
good state, the value of $A$ becomes 
much higher than that of $B$, which makes it slow to switch. As
both actions are available in most rounds, this leads to a low
probability of survival for the value strategy.

Having more than two states becomes more useful when 
playing $B$ is bad in the good environment (i.e.,
$P_{GB}<0.5$) and the signal from playing $A$ is not very strong
(i.e., $P_{SA}$ and $P_{GA}$ are not too far apart).
Note however, that these condition
are outside the scope of parameters allowed by the model specified by
McNamara et al.~\citeyear{MTH12} 
(as $P_{GB}<P_{SB}$ if we want a good payoff for $B$ in the sparse environment).


\section{Discussion}\label{sec:learning}

While we show that the automaton strategy is better than the value
strategy in many scenarios, as we discussed before, the value strategy
(or, more generally, state-dependent strategies), seem to
be what animals actually use in the scenarios studied in previous
papers. We 
now discuss some possible explanations for this. 

\subsection{The learning challenge}
One advantage of the value strategy is that it does not need
to ``understand'' that $A$ is the ``risky'' action and $B$ is the
``safe'' action.  It just calculates a value for each strategy and
plays the strategy with the higher value.  By way of contrast, there is
a sense in which the safe strategy and the automaton strategy need to
understand beforehand that $B$ is safe.
As long as roles are reasonably stable
(over evolutionary timescales), this is not a problem; 
evolutionary pressures would result in animals effectively learning 
which actions are safe and which are risky.  And if
roles change
over time, then again we would expect mutations to 
survive that switched
the role of $A$ and $B$, provided that the change
is sufficiently slow.


What happens in new environments where the animal does not initially
know which actions are safe 
or in environments where the roles of ``safe'' and ``risky'' actions
change relatively frequently?
It is easy to extend the automaton strategy by adding a
``front end'', so to speak, that keeps learning which action is safe and
which is risky.  For example, the agent can keep track of the payoffs in the
last $m$ rounds (for a small value of $m$, say 10-15) to see if they
are in line with the current choices of safe and risky actions; if
not, the choices of ``safe'' and ``risky'' can be switched.
If the payoff difference between the safe and risky strategies
is relatively large (in at least one state), then this can be learned
quickly; if it not so large, then it does not matter so much which
action is performed. 

It also worth noting that the value strategy also has some
parameters ($\beta$ and $k$) that might depend on the environment and
require some learning process. 
These observations suggest that 
learning by itself is not the explanation for the
usage of the value strategy by animals. 

\subsection{A hybrid strategy}
Another possible explanation is that animals use a hybrid
strategy, combining features of both the value strategy and the
automaton strategy.%
\footnote{We thank a reviewer of the paper for suggesting the use of
  the hybrid strategy.}
The value strategy seems to do well in  learning the
value of actions in a new environment and reacting to changes in the
value of actions themselves. However, the automaton strategy is better
at tracking the current environment and reacting to somewhat periodic
changes in the environment. 
%
The use of a hybrid strategy is supported by the fact that the
animal's preference for food sources 
presented during tough times is ephemeral; as soon as animals
experience both sources in the same state, they start switching
preference to the option with higher resource payoff [Alex Kacelnik,
  private communication, 2016].  Thus, when the environment 
stabilizes, animals might be switching from a state-dependent strategy
to using the ``safe" strategy.
Of course, more research is needed to see if this is what actually happens.

\section{Conclusion}\label{sec:discussion}


Our results show that some very simple strategies seem to
consistently outperform the value
strategy. This gap grows as the task of surviving becomes more
challenging (either because ``winter" is longer or the
rewards are not as high). 
This at least suggests that the model considered by McNamara et 
al.~\citeyear{MTH12} is not sufficient to explain the evolution of
the value strategy in animals.
McNamara et al.~claim that
``[w]hen  an animal lacks knowledge of the environment it faces, it may
be adaptive for the animal to base its decision upon approximate
cues.''  We are very sympathetic to this claim, although we have
argued that it may be more useful for those cues to include recent evidence
about rewards, not just the internal food state, at least for the
parameter settings considered by McNamara et al.  
It would be interesting to understand better how the interaction of
cues is used by animals.
\medskip
\bibliographystyle{chicago}
\bibliography{joe,lior}

\end{document}